\documentclass[aps,prb,twocolumn,superscriptaddress]{revtex4-2}
\usepackage{graphicx} 

\usepackage{xcolor}

\newcommand{\muB}{\mu_\mathrm{B}}
\newcommand{\up}{\uparrow}
\newcommand{\dn}{\downarrow}

\usepackage{xr}
\begin{document}

\title{Giant spatial anisotropy of magnon lifetime in altermagnets}
\author{A. T. Costa}
\affiliation{International Iberian Nanotechnology Laboratory (INL), Av. Mestre Jos\'e Veiga, 4715-330 Braga, Portugal}
\author{J. C. G. Henriques}
\affiliation{International Iberian Nanotechnology Laboratory (INL), Av. Mestre Jos\'e Veiga, 4715-330 Braga, Portugal}
\affiliation{Universidade de Santiago de Compostela, 15782 Santiago de Compostela, Spain}
\author{J. Fern\'andez-Rossier}
\altaffiliation[On permanent leave from ]{Departamento de F\'isica Aplicada, Universidad de Alicante, 03690 San Vicente del Raspeig, Spain.}
\affiliation{International Iberian Nanotechnology Laboratory (INL), Av. Mestre Jos\'e Veiga, 4715-330 Braga, Portugal}
\date{\today}

\begin{abstract}
    Altermagnets are a new class of magnetic materials with zero net magnetization (like 
    antiferromagnets) but spin-split electronic bands (like ferromagnets) over a fraction of 
    reciprocal space. As in antiferromagnets, magnons in altermagnets  come in two flavours, that  
     either add one or remove one unit of spin to the $S=0$ ground state. However, in altermagnets these two magnon modes  are non-degenerate along some directions in reciprocal space. 
    Here we show that the lifetime of altermagnetic magnons has a very strong dependence on both flavour and direction. Strikingly, coupling to Stoner modes  leads to a complete suppression of magnon propagation along selected spatial directions. This giant anisotropy will impact electronic, spin, and energy transport properties and may be exploited in spintronic applications.
\end{abstract}

\maketitle
The recent recognition of altermagnets as a new class of magnetic materials~\cite{RashbaZunger2020,EmergingLandscape2022,Smejkal2022}, originally predicted by Pekar and Rashba in 1964~\cite{Pekar1964}, 
has been a very exciting development for both condensed matter and materials
physics. In a static configuration, altermagnets camouflage very well as antiferromagnets; however,  
when you look under the hood the disguise is given away by the spin-polarized
electronic bands. It is their dynamics, however, that reveal their true colors \cite{Maier2023,Sarkar2024}. 
To understand the dynamical properties of a magnetic system it is essential to 
look at its elementary spin excitations, or magnons~\cite{Takahashi2013}. 

A magnon in a ferromagnetic 
solid is usually associated to processes by which the total magnetization of 
the sample is lowered by the equivalent of
a quantum of angular momentum, $\hbar$, and associated with the spin-lowering
operator $S^-$. We thus say that a ferromagnetic magnon carries spin $S^z=-1$. 
In terms of elementary electronic processes, generating a magnon consists in
promoting an electron from the majority spin band ($\up$) to the minority 
spin band ($\dn$), and is associated with the operator $a^\dagger_\dn a_\up$.
By virtue of electron-electron interactions, the electron and the hole involved
in this process form a bound state, whose energy depends on the net crystal 
momentum of the pair. 

In antiferromagnets,
magnons can have either $S^z=-1$ or $S^z=1$, associated with lowering the spin of
the $\up$ sublattice or raising the spin of the $\dn$ sublattice. Due to the complete
equivalence between the two spin directions, the two kinds of antiferromagnetic 
magnons ($S^z=\pm 1$) have identical energies~\cite{Rezende_2019}. 
On the other hand, it has been noted \cite{EmergingLandscape2022,MagnonsRuO2PRL} 
that magnons in altermagnets have unique features when compared to their 
antiferromagnetic counterparts. The most noticeable difference it that $S^z=-1$ and $S^z=1$ magnons
have distinct energies along certain directions in the reciprocal space, the same direction
associated with the spin-split electronic bands. 

In metallic magnets, magnons have finite lifetimes, due to the fact that they 
can decay into uncorrelated electron-hole pairs, also known as a Stoner excitations~\cite{doniach1998green,Barbosa2001}.
The decay probability (hence the inverse of the magnon lifetime) is proportional 
to the spectral density associated with the Stoner excitations, which usually
increases monotonically with energy for a fixed wavevector. Thus, magnon
lifetimes typically decrease monotonically as the magnon energy increases~\cite{lifetimeCoSW}.

It has been assumed hitherto \cite{MagnonsRuO2PRL} 
that, due to the distinct energies of $S^z=\pm 1$ magnons in altermagnets, 
their lifetimes would also be different, in an almost trivial manner.
Other works have looked into the effects of magnon-magnon interactions 
on magnon lifetimes, a mechanism that is supposed to be relevant for insulating magnets.~\cite{garciagaitan2024magnon} Apart from that, very little attention 
has been payed to the lifetime of magnons in altermagnets, and most theoretical
approaches employ spin-only models in their description \cite{MagnonsRuO2PRL,sodequist2024twodimensional,consoli2024su3,hodt2023spin}.

Here we show that Stoner damping in metallic and 
slightly doped altermagnets has highly non-trivial consequences. 
Specifically, the combination between the peculiar symmetry of the 
altermagnet and the damping by Stoner excitations makes magnons in 
itinerant altermagnets completely distinct from their antiferro- and 
ferromagnetic counterparts. The \textit{altermagnons} acquire a strong 
frequency- and spin-dependent directionality, which can potentially be 
exploited as a resource in spintronics devices~\cite{ElKanj2023}. 

\begin{figure}
    \centering
    \includegraphics[width=\columnwidth]{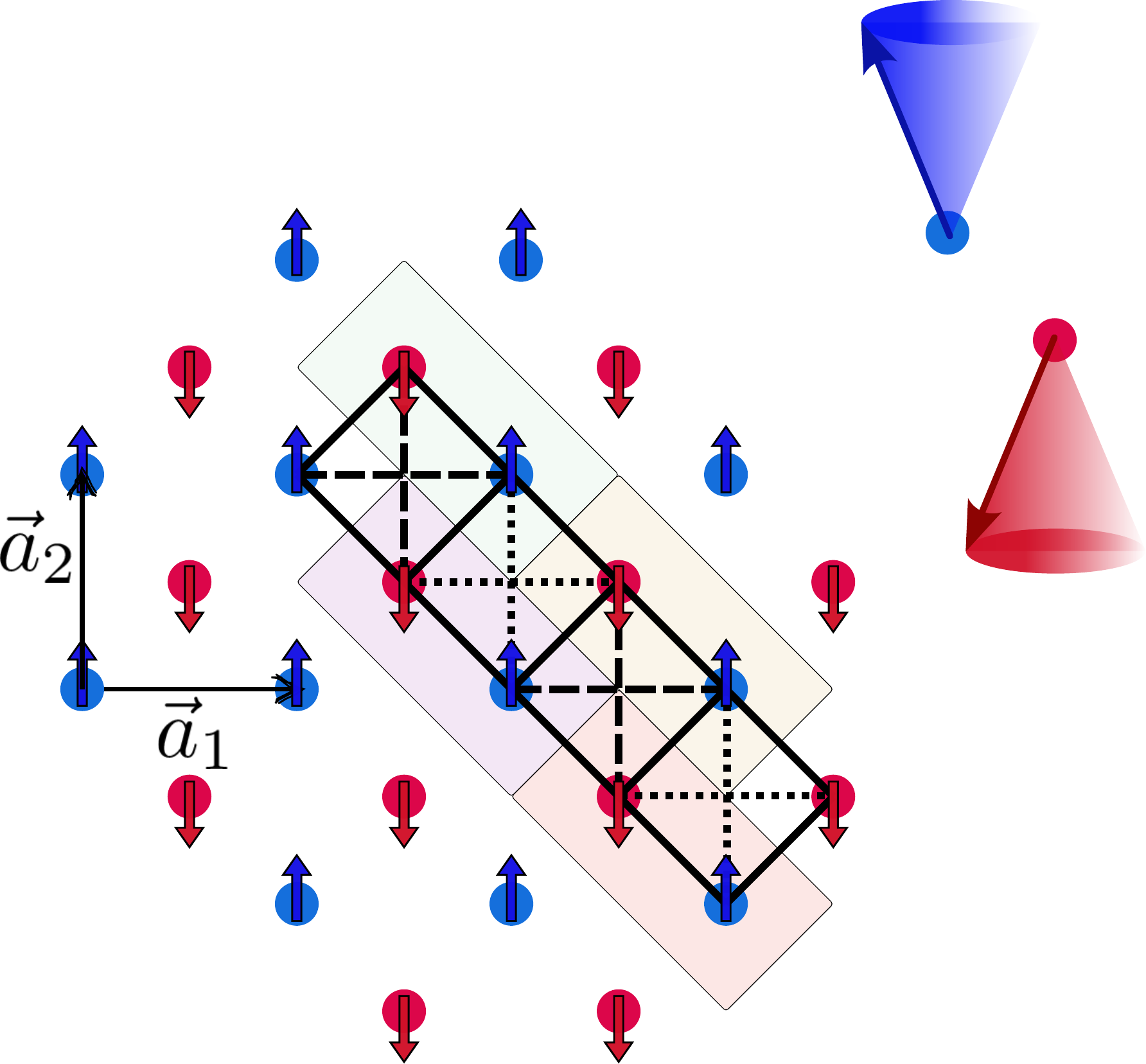}
    \caption{Schematic representation of the model altermagnet on a square lattice defined by primitive vectors $\vec{a}_1 $ and $\vec{a}_2$, with $|\vec{a}_1|=|\vec{a}_1|=a$. The solid line connecting blue and red sites represents the nearest neighbor hopping $\tau$. Dashed and dotted lines represent the alternating second neighbor hoppings $\tau'(1\pm\delta)$.}
    \label{fig:model}
\end{figure}

\textsl{Model and mean-field ground state:} We model the electronic structure of altermagnets using a Hamiltonian proposed in ref.~\cite{das2023realizing}, which is essentially a Hubbard model with an especially chosen hopping 
structure that realises an altermagnetic symmetry,
\begin{equation}
    H = \sum_{ll'}\sum_{\mu\mu'}\sum_\sigma \tau_{ll'}^{\mu\mu'}c^\dagger_{l\mu\sigma}c_{l'\mu'\sigma} + U\sum_{l,\mu} n_{l\mu\up}n_{l\mu\dn},
    \label{eq:hamilt}
\end{equation}
where $n_{l\mu\sigma}\equiv c^\dagger_{l\mu\sigma}c_{l\mu\sigma}$, $l$ and $l'$ label unit cells, $\mu$ and $\mu'$ label sublattices ($A$ or $B$) and $\sigma$ labels the spin projection along the $z$ axis. The hopping matrix $\tau^{\mu\mu'}_{ll'}$ is described in the caption of Fig.~\ref{fig:model}. The intra-atomic interaction parameter $U$ can be chosen to place the system in either the metallic or insulating altermagnetic phase; for the value of diagonal hopping we adopted in this work, $2\tau\lesssim U\lesssim 3\tau$ yields a metallic altermagnetic phase, whereas $U\gtrsim 3\tau$ produces the insulating altermagnetic phase. The complete mean-field phase diagram of this model has been explored in Ref.~\cite{das2023realizing}. Here we will choose two representative points, one in the insulating and one in the metallic region, and study the elementary spin excitations above their respective mean-field ground states. The mean-field approximation we employ amounts to the following replacement,
\begin{eqnarray}
    U\sum_{l,\mu}n_{l\mu\up} n_{l\mu\dn}\longrightarrow\nonumber\\
    \frac{U}{2}\sum_{l,\mu}\left[  (\bar{n}_{\mu l}+\bar{m}_{\mu l})n_{l\mu\dn} + 
   (\bar{n}_{\mu l}-\bar{m}_{\mu l})n_{l\mu\up}  \right],
\end{eqnarray}
with $\bar{n}_{\mu l}\equiv \langle n_{l\mu\up}\rangle + \langle n_{l\mu\dn}\rangle$
and $\bar{m}_{\mu l}\equiv \langle n_{l\mu\up}\rangle - \langle n_{l\mu\dn}\rangle$,
plus a constant term that can be safely ignored. The average occupancies $\bar{n}_{\mu l}$ and magnetic moments $\bar{m}_{\mu l}$ are determined self-consistently.

We obtain the magnon spectrum of altermagnets by studying the transverse spin susceptibilities,
\begin{equation}
    \chi^{+-}_{\mu\nu}(\vec{r}_{l'}-\vec{r}_{l},t)\equiv -i\theta(t)\left\langle\left[S^+_{l\mu}(t),S^-_{l\nu}(0)\right]\right\rangle
    \label{eq:chipm}
\end{equation}
and
\begin{equation}
    \chi^{-+}_{\mu\nu}(\vec{r}_{l'}-\vec{r}_{l},t)\equiv -i\theta(t)\left\langle\left[S^-_{l\mu}(t),S^+_{l\nu}(0)\right]\right\rangle ,
    \label{eq:chimp}
\end{equation}
where $t$ is the time, $S^-_{l\mu}\equiv c^\dagger_{l\mu\dn}c_{l\mu\up}$, ($S^+_{l\mu} = (S^-_{l\mu})^\dagger$)) is the operator that creates a spin excitation with $S^z=-1$ ($S^z=1$) at cell $l$ in the sublattice $\mu$, $\vec{r}_l$ is the position of unit cell $l$, and $\theta(t)$ is the Heaviside unit step function.
These two-time correlation functions cannot be computed exactly for an interacting model such as
the one defined in Eq.~\ref{eq:hamilt}; the simplest approach that can describe
magnons is the so-called random phase approximation (RPA), in which the interaction is
taken into account, to all orders in perturbation theory, between the electron and the hole
that form the spin-flip excitation~\cite{doniach1998green}. The RPA relates the interacting susceptibilities $\chi^\perp$ to
the mean-field susceptibilities $\bar{\chi}^\perp$, which are the same Green functions defined in Eqs.~\ref{eq:chipm} and~\ref{eq:chimp}, with the thermal average $\langle\cdot\rangle$ evaluated for the mean-field configuration. 
For the model considered here, after Fourier transforming both in time and position, the RPA equations are
\begin{eqnarray}
    \chi^{+-}_{\mu\nu}({\cal Q}) = \bar{\chi}^{+-}_{\mu\nu}({\cal Q}) -
    U\sum_{\xi}\bar{\chi}^{+-}_{\mu\xi}({\cal Q})\chi^{+-}_{\xi\nu}({\cal Q}),
\end{eqnarray}
where $\mathcal{Q}\equiv(\vec{q},\hbar\Omega)$. We obtain an analogous expression for $\chi^{-+}$.
The spectral density associated with magnons, projected on sublattice $\mu$, is given by
\begin{equation}
    \rho^\perp_\mu(\mathcal{Q}) = -\frac{1}{\pi}\mathrm{Im}\chi^{\perp}_{\mu\mu}(\mathcal{Q})
\end{equation}
where $\perp$ can be either $+-$ or $-+$, denoting the transversal character of these response functions with respect to the equilibrium staggered magnetization (Néel vector). Magnon energies $\hbar\Omega(\vec{q})$ are associated with the positions of the peaks of $\rho^{+-}$ (for the $S^z=-1$ magnons) or $\rho^{-+}$ (for the $S^z=1$ magnons), at fixed wave-vector $\vec{q}$. Analogously, magnon lifetimes are defined as the inverse of the full width at half-maximum of the magnon peaks.

\textsl{Mean-field results:} An insulating altermagnetic state can be obtained by choosing $U\gtrsim 3\tau$; however, for $3\tau\lesssim U\lesssim 10\tau$ the mean-field configuration
belongs to an intermediate coupling regime, for which the spin dynamics can not yet be properly described by a spin-only (Heisenberg-like) model~\cite{suppl}. Thus, to benchmark our
fermionic model against a spin model, we chose $U=10\tau$, together with the hopping values $\tau'=0.17\tau$ and $\delta=0.83\tau$. The self-consistent mean-field solution gives the bands shown in Fig.~\ref{fig:el_bands_1} of the supplementary material, with a staggered magnetic moment $m_A-m_B=1.86\muB$ per unit cell. For the reciprocal space path we plotted in Fig.~1 of the supplementary materials, the spin splitting is zero only along the line $q_y=q_x$. Along the line $q_y=\frac{\pi}{a}-q_x$ there is the characteristic crossing between the $\up$ and $\dn$ spin bands, associated with the altermagnetic symmetry.

The metallic altermagnetic state can be obtained either by tweaking the hopping parameters, as shown in ref.~\cite{das2023realizing}, or by reducing the Hubbard parameter $U$. We chose the latter option to minimize
the differences between the shapes of the electronic bands in the metallic and insulating states. 
By setting $\tau'=0.17\tau$, $\delta=0.83\tau$ and $U=2.5\tau$ we obtain the metallic altermagnetic bands shown in Fig.~1 of the supplementary material, 
with a staggered magnetic moment $m_A-m_B=0.74\muB$ per unit cell.

\begin{figure}
    \centering
    \includegraphics[width=\columnwidth]{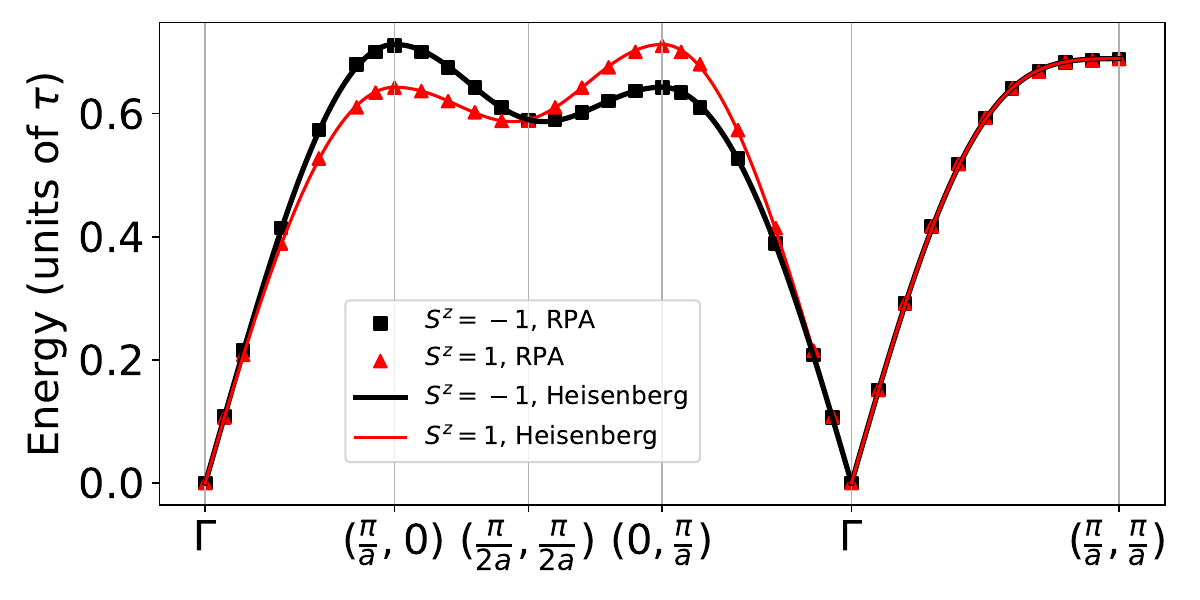}
    \caption{Dispersion relation for magnons in an insulating altermagnet in the strong coupling regime ($U=10\tau$). The Heisenberg model used to fit the RPA energies includes up to third-neighbor exchange.}
    \label{fig:sw_dispersion_insulating}
\end{figure}

\textsl{Altermagnons:} To benchmark our methodology, we first 
analyze the spin excitations of the insulating altermagnet in
the strong coupling limit ($U=10\tau$), for which the spin model results should be valid \cite{BrokenSymmetry2023,consoli2024su3}.
By scanning the spectral densities $\rho^{+-}$ and $\rho^{-+}$ in the $(\hbar\Omega,\vec{q})$ 
space we obtain the dispersion relations for $S^z=-1$ altermagnons ($+-$) and for $S^z=1$ altermagnons ($-+$), shown in
Fig.~\ref{fig:sw_dispersion_insulating}. The energy splitting between the two polarizations, 
one of the hallmarks of altermagnetism, is clearly seen along high-symmetry directions 
in the Brillouin zone. We also show the dispersion relation for (linearized) Holstein-Primakoff 
magnons, extracted from a Heisenberg model for the altermagnet, including up to third-neighbor 
exchange. As expected, the agreement with the RPA treatment of the fermionic model is very good in this case~\footnote{This is contrast with the insulating intermediate coupling case, 
for which the spin model fails. See supplemental material, section~\ref{sec:intermediate} \cite{suppl}.}.

Along specific lines within the BZ we observe a behavior analogous to the ``band inversion'' associated 
with topologically non-trivial electronic bands. For instance, along the reciprocal space path going 
from $(\frac{\pi}{a},0)$ to $(0,\frac{\pi}{a})$ there is a crossing between the $S^z=-1$ and the 
$S^z=1$ altermagnon branches. In the presence of spin-orbit coupling a gap may appear at the crossing 
point $(\frac{\pi}{2a},\frac{\pi}{2a})$, possibly accompanied by a finite Berry curvature. This 
crossing is also associated with the peculiar directional behavior of altermagnetic magnons. If we 
focus on altermagnons with one $S^z$ value we see that the energy at the $(\frac{\pi}{a},0)$ point in 
reciprocal space (thus, propagating along the $x$ direction in real space with wavelength $\lambda=2a$) 
is 40\% different from that of a magnon with the same wavelength propagating along the $y$ direction. 
This is illustrated in fig.~\ref{fig:SM_radial_insulator} of the supplementary material~\cite{suppl}, where we plot the altermagnons spectral 
densities as a function of propagation direction, for a fixed wavelength. Combined with the fact that, 
for sufficiently small wavelengths (typically smaller than $\sim 5a$) magnons with a well-defined $S^z$ are strongly sublattice-polarized, this feature may be exploited to guide magnons in spintronics devices. 

\begin{figure}
    \centering
       \includegraphics[width=\columnwidth]{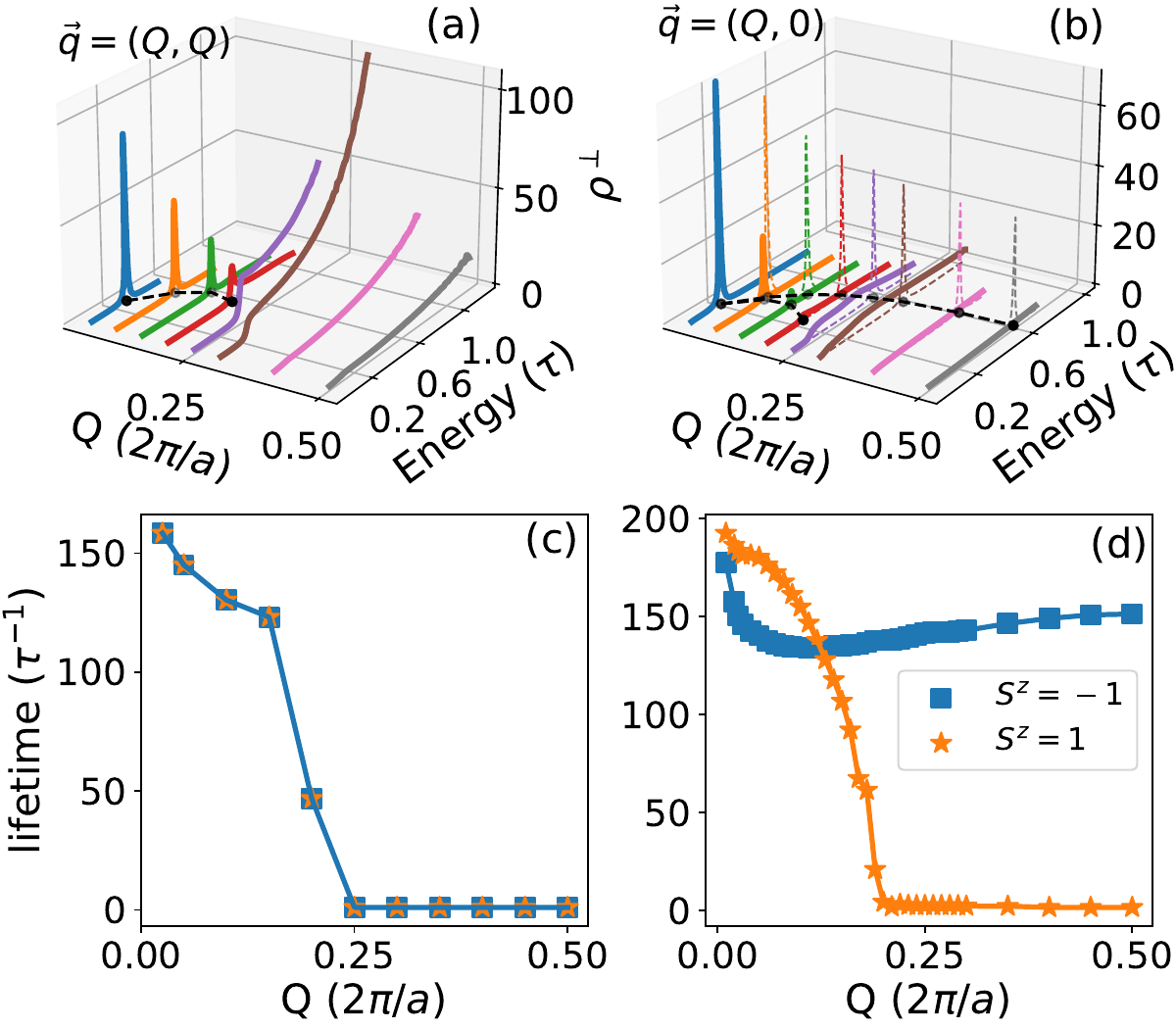}
    \caption{
    Top: spin excitation spectral densities in the metallic phase ($U=2.5\tau$),  
    along $\vec{q}=(Q,Q)$ (a) and $\vec{q}=(Q,0)$ (b), as a function of energy, 
    for selected wave numbers. To improve visualization, the spectral density has 
    been multiplied by 100 for the three largest wavenumbers ($Q=0.3$, 0.4 and 0.5), 
    by 50 for $Q=0.25$ and by 5 for $Q=0.2$. In (b), solid lines correspond to $\rho^{-+}$, 
    associated with the $S^z=1$ spin excitations, and dashed lines correspond to $\rho^{+-}$, 
    associated with the $S^z=-1$ spin excitations. Bottom: Lifetimes of the metallic 
    altermagnons ($U=2.5\tau$) propagating along the $\vec{q}=(Q,Q)$ 
    (c) and $\vec{q}=(Q,0)$ (d), as a function of wave number, 
    for $S^z=-1$ (squares) and $S^z=1$ (stars) spin excitations.}    
    \label{fig:3}
\end{figure}

We now turn our attention to the behaviour of magnons in itinerant altermagnets.
In contrast to the insulating case, it can be expected that their  lifetime is limited by   Stoner damping~\cite{doniach1998green,Barbosa2001,Costa06SWdamping}. Magnons with energies exceeding 
single-particle spin-flip excitations (also known as Stoner excitations), 
can decay into the Stoner continuum~\cite{doniach1998green}. The magnon lifetime is 
inversely proportional to the density of Stoner modes, which in given by the imaginary part of the 
mean-field transverse susceptibility $\bar{\chi}^\perp$. 
The effect of damping for a conducting altermagnet ($U=2.5\tau$) is seen in the 
evolution of the spectral weight of spin excitations, shown along two different directions, 
$(Q,Q)$ and $(Q,0)$ in Figure \ref{fig:3}a,b. For low energy, the spectral density has well defined peaks, 
whose position gives the magnon energy and the inverse of its linewidth gives the magnon lifetime. 
As the energies are increased, the peaks get broader and, above some energy threshold, 
they vanish into a continuum.  Along the $(Q,Q)$ direction, both $S_z=\pm 1$ excitations have the same 
spectral weight (fig.~\ref{fig:3}a). In contrast,
along the $(Q,0)$ direction (fig.~\ref{fig:3}b), the $S^z=-1$ spin excitations 
have lorentzian spectral densities with relatively small linewidth in the whole wave number
range, whereas the spectral density associated with the $S^z=1$ spin excitations has
a behavior similar to the $(Q,Q)$ case. We thus find that, for  itinerant altermagnets,  
magnons with a given $S^z$ are only well defined along certain directions.

\begin{figure}
    \centering   
     \includegraphics[width=\columnwidth]{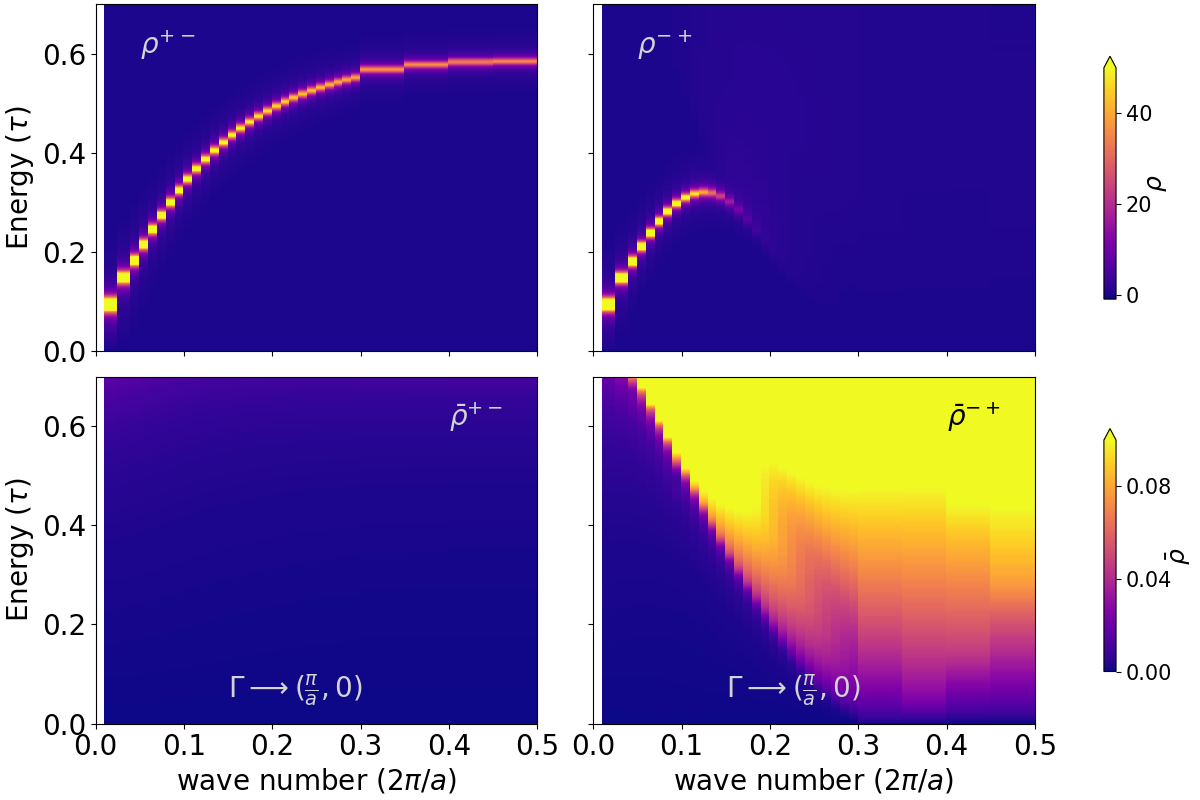}
    \caption{Top: Spectral densities for $S^z=-1$ ($\rho^{+-}$, left) and $S^z=1$ ($\rho^{-+}$, right) 
    metallic altermagnons ($U=2.5\tau$) propagating along the $x$ direction, as a function of wave number and  energy.
    Bottom: Spectral densities for $S^z=-1$ ($\bar{\rho}^{+-}$, left) and $S^z=1$ ($\bar{\rho}^{-+}$, right) Stoner excitations (single-particle spin flips) propagating along the $x$ direction, as a function of wave number and  energy.}
    \label{fig:rhoQx}
\end{figure}

To make the connection between magnon lifetimes and density of Stoner modes, it
is useful to plot both magnons' and Stoner excitations' spectral densities as color-coded
functions of energy and wave number, shown in fig.~\ref{fig:rhoQx}. By following the bright 
spots in the top left panel, it is possible to trace dispersion relations for the $S^z=-1$ magnons, 
in analogy to the insulating case. For the $S^z=-1$ magnon, the bright spots disappear around
$Q\sim \frac{2\pi}{5a}$. This can be correlated with the boundaries of the Stoner continuum
for $S^z=1$ spin excitations, plotted in the bottom right panel. In contrast, the density of
$S^z=-1$ Stoner modes is uniformly small over the whole wave number and energy ranges where 
$S^z=-1$ magnons exist. A discussion of the origin of the density of Stoner modes in terms
of the geometry of the spin-polarized Fermi contours of the metallic altermagnet is presented in 
the supplemental material~\cite{suppl}.

\begin{figure}
    \centering
    \includegraphics[width=\columnwidth]{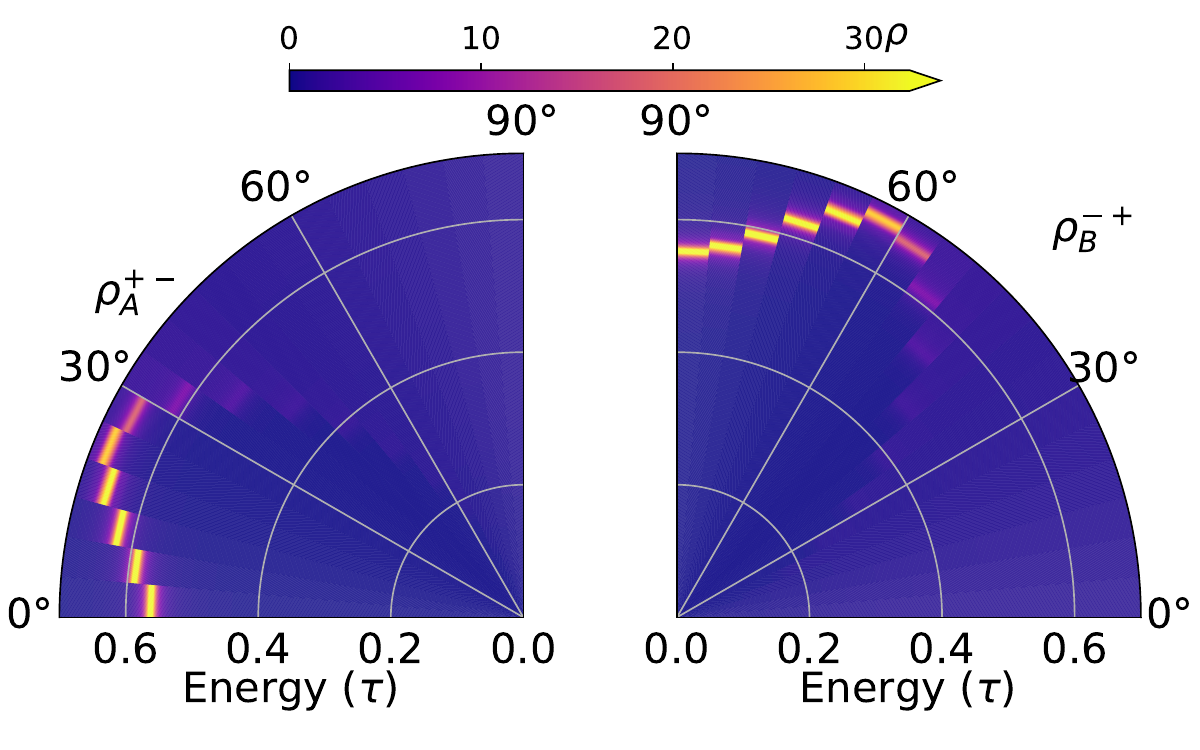}
    \includegraphics[width=\columnwidth]{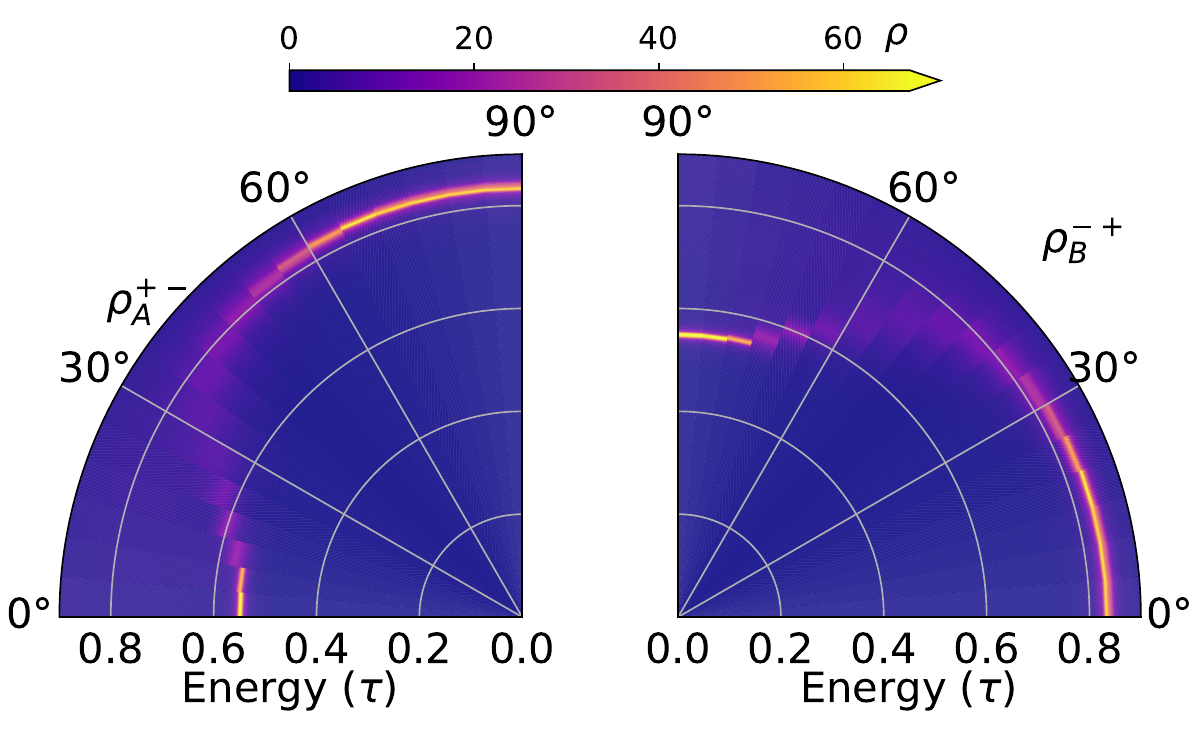}
    \caption{Altermagnon spectral densities as functions of propagation angle, for a fixed wavelength ($\frac{10a}{3}$). The radial variable represents energy (in units of the nearest-neighbor hopping $\tau$). $\rho^{+-}_A$ corresponds to $S^z=-1$ magnons, $\rho^{-+}_B$ corresponds to $S^z=1$ magnons. Top panels: metallic phase ($U=2.5\tau$);  bottom panels: doped insulating phase ($U=3.5\tau$, excess 0.1 electrons per unit cell).}
    \label{fig:radialSD}
\end{figure}

The giant magnon-lifetime anisotropy is better seen in
a color-coded polar plot of the magnon spectral density, for a fixed 
wavelength. The angular variable indicates the propagation direction, and the
radial variable is the magnon energy. In fig.~\ref{fig:radialSD} we show such a plot
for $\lambda=\frac{10a}{3}$ (wave number $Q=\frac{3\pi}{5a}$). The top-left panel shows 
the spectral density $\rho^{+-}_A$ for $S^z=-1$ altermagnons, projected on sublattice 
$A$, and the top-right panel displays the equivalent quantity for sublattice $B$ ($\rho^{+-}_B$). 
It is clear that $S^z=-1$ altermagnons are strongly suppressed for angles $\gtrsim 30^\circ$, 
and the $S^z=1$ altermagnons for angles $\lesssim 60^\circ$. 
Such strong directionality is rarely seen for quasiparticles and elementary excitations, 
and is potentially very useful for applications, especially when one considers 
the fact that altermagnons of wavelengths $\lambda\lesssim 4a$ live preferentially in one 
of the sublattices. Thus, it is in principle possible to excite altermagnons along specific 
directions by choosing their excitation frequency and the sublattice to excite. Selectively 
addressing the sublattice may be challenging in systems where spin sublattices have atomic 
size, but not so much in synthetic magnets, where spin sublattices are associated with 
molecules containing tens of atoms~\cite{BrokenSymmetry2023,liu2024twisted}.

We have also considered the case of a doped insulating altermagnet, by choosing $U=3.5\tau$ and imposing an electronic occupation of 1.05 electrons per atomic site. 
In this case the anisotropic suppression of altermagnons is observed for propagation 
angles $30^\circ\lesssim \theta \lesssim 75^\circ$, as shown in the bottom panels of 
fig.~\ref{fig:radialSD}. Thus, whenever it is possible to dope an insulating altermagnet 
electrostatically, it is in principle also possible to control electrostatically the propagation 
direction of altermagnons.

The effects of a giant spatial anisotropy in magnon lifetimes are likely to be noticed 
on several transport coefficients of metallic altermagnets~\cite{SpintronicsRMP2004}. 
Electronic transport is expected to be impacted by electron-magnon scattering, especially at
low temperatures. 
Moreover, with current high-resolution spin-polarized electron energy loss 
spectroscopy~\cite{ZakeriHRSPEELS,ZakeriFeSe} it should be possible to probe experimentally
the lifetime anisotropy predicted by our theoretical analysis.

In conclusion, we have studied the intrinsic damping of magnons in altermagnets. These collective modes come with two values of $S_z=\pm 1$. Contrary to their counterparts in ferro- and antiferromagnets, we find  a giant spatial anisotropy of magnon lifetimes in itinerant altermagnets. We find that, for a given direction, only magnons with a given sign of $S_z$ survive without melting due to Stoner damping. The ultimate reason for this unique behaviour underlies on the existence of spin-polarized Fermi surface pockets that characterizes altermagnets. Therefore, we expect our predictions are generic of all itinerant altermagnets, rather than model specific and will have to be considered in future altermagnonic applications.

\begin{acknowledgments}
A.T.C. acknowledges fruitful discussions with D. L. R. Santos.
The authors acknowledge financial support from 
 FCT (Grant No. PTDC/FIS-MAC/2045/2021),
SNF Sinergia (Grant Pimag,  CRSII5\_205987)
 the European Union (Grant FUNLAYERS
- 101079184).
J.F.-R. acknowledges financial funding from 
Generalitat Valenciana (Prometeo2021/017
and MFA/2022/045),
Spanish Government through PID2022-141712NB-C22, 
and the Advanced Materials programme  supported by MCIN with funding from European Union NextGenerationEU (PRTR-C17.I1) and by Generalitat Valenciana (MFA/2022/045).
\end{acknowledgments}


\bibliographystyle{apsrev4-2}
\bibliography{references}

\appendix







\section{Mean-field electronic structure}
\begin{figure*}
    \centering
    \includegraphics[width=\columnwidth]{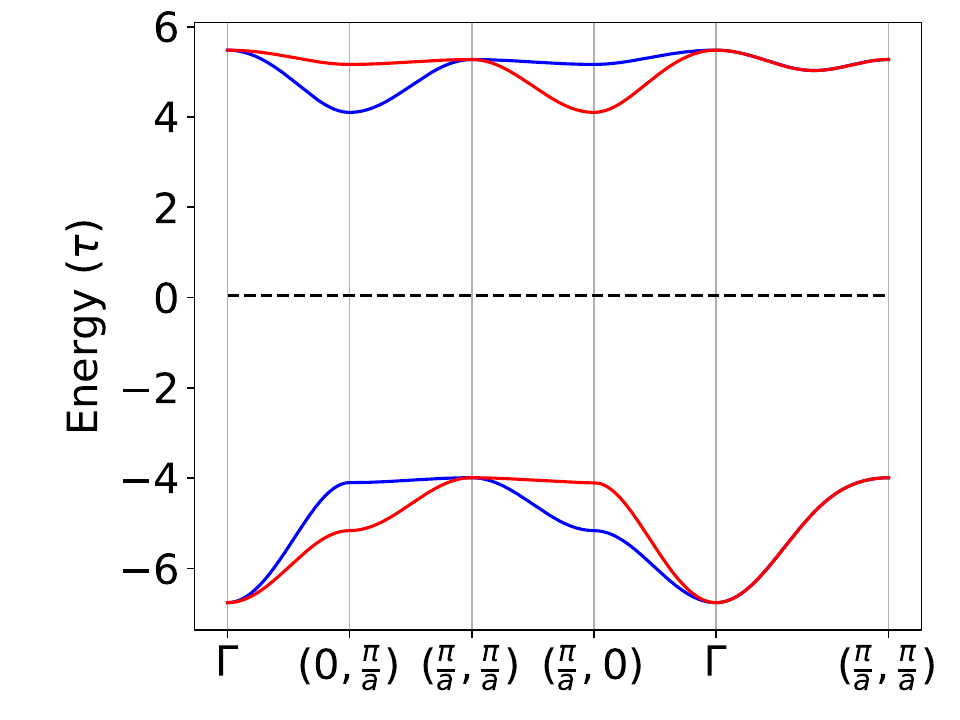} 
    \includegraphics[width=\columnwidth]{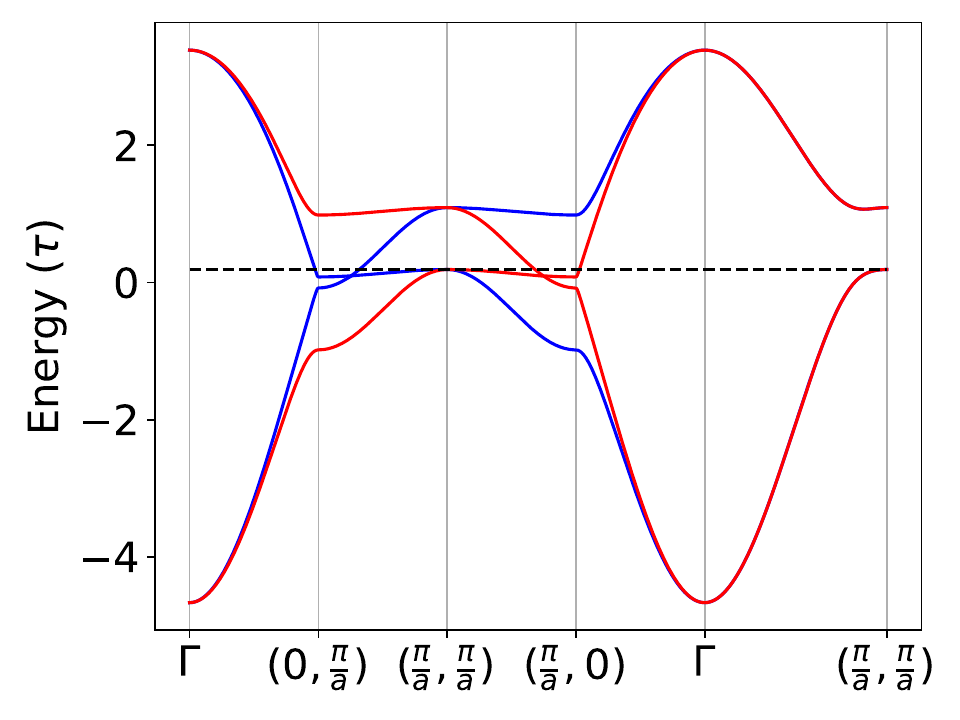}
    \caption{Electron energy bands for the strong-coupling insulating (left panel, $U=10\tau$) and metallic (right panel, $U=2.5\tau$) mean-field ground state configuration of the altermagnet Hamiltonian (eq.~\ref{eq:hamilt} of the main text), with $\tau'=0.16\tau$ and $\delta=0.83$. Red and blue lines represent $\up$ and $\dn$ spin sub-bands. The black dashed line marks the Fermi energy. }
    \label{fig:el_bands_1}
\end{figure*}

\begin{figure*}
    \centering
     \includegraphics[width=\columnwidth]{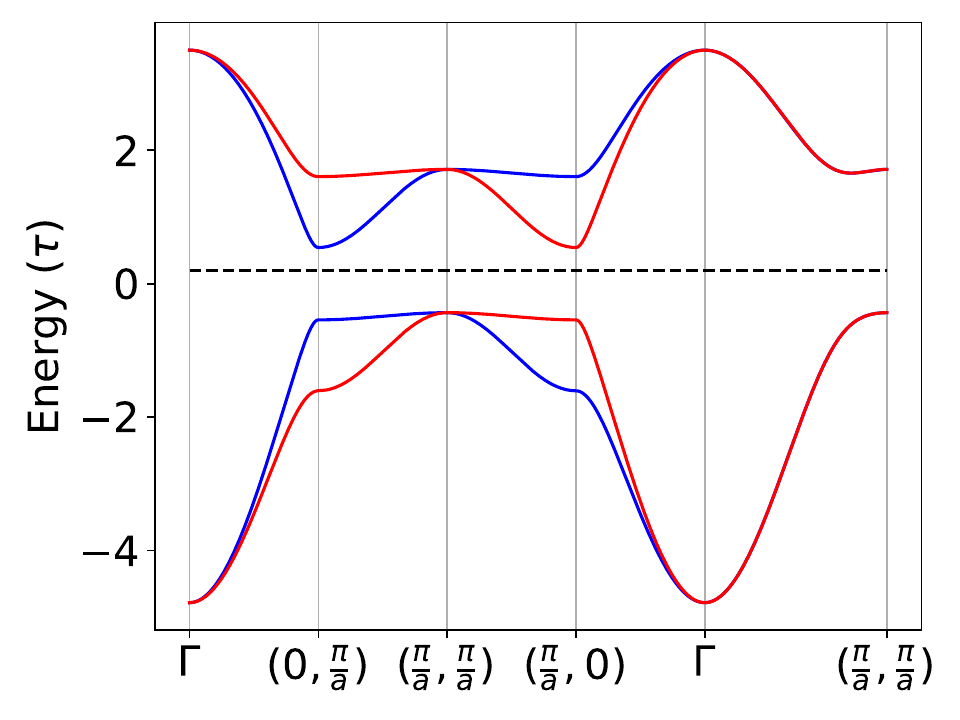} 
     \includegraphics[width=\columnwidth]{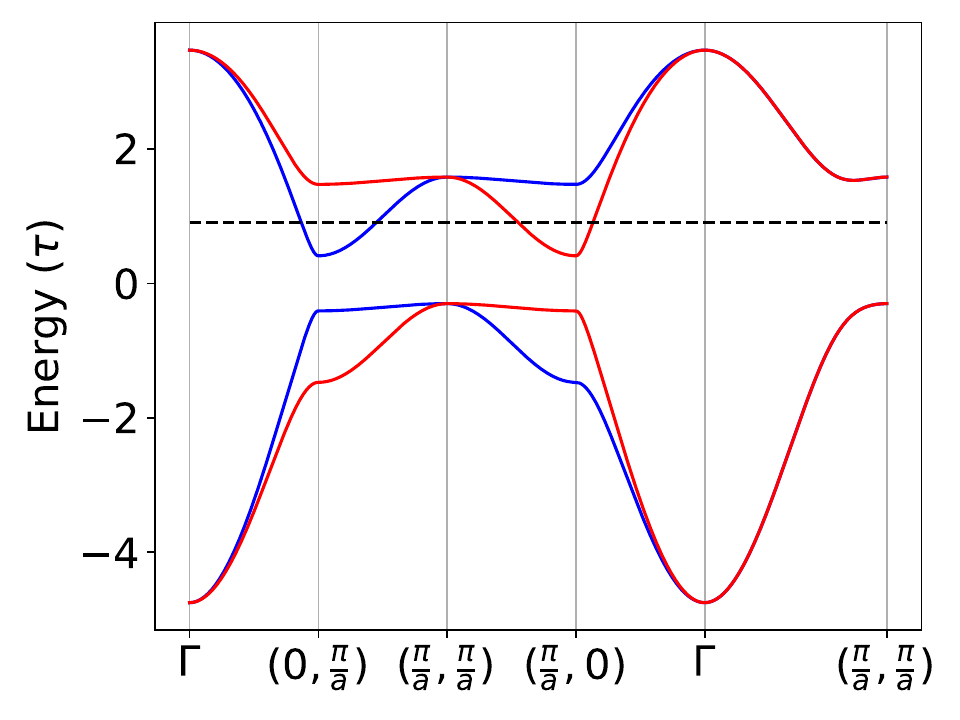}
    \caption{Electron energy bands for the mean-field ground state configuration of the altermagnet Hamiltonian (eq.~\ref{eq:hamilt} of the main text) in the insulating intermediate coupling regime ($U=3.5\tau$) at half-filling (left panel) and away from half-filling (1.05 electrons per lattice site, right panel). The values for the hopping parameters are $\tau'=0.16\tau$, $\delta=0.83$. Red and blue lines represent $\up$ and $\dn$ spin sub-bands. The black dashed line marks the Fermi energy. }
    \label{fig:el_bands_2}
\end{figure*}

We present the electronic bands corresponding to the mean-field configurations considered in the letter:
strong-coupling insulating ($U=10\tau$, fig.~\ref{fig:el_bands_1}, left panel), metallic ($U=2.5\tau$, fig~\ref{fig:el_bands_1}, right panel), and slightly doped insulating ($U=3.5\tau$, fig.~\ref{fig:el_bands_2}, right panel). Both metallic and insulating phases have half-filled bands (one electron per lattice site), whereas the doped phase has 1.05 electrons per lattice site. Table~\ref{tab:my_label} shows the values of the Hamiltonian parameters associated with the different phases, as well as the mean-field staggered magnetic moment per unit cell.
We also show the intermediate-coupling insulating case ($U=3.5\tau$, fig.~\ref{fig:el_bands_2}, left panel).

\begin{table}[h]
    \centering
    \begin{tabular}{c|c|c|c|c}
        &  $\tau'$ & $\delta$ & $U$ & $|m_\up - m_\dn| (\mu_\mathrm{B})$ \\ \hline
       Insulating (strong coupling)  & 0.16 & 0.83 & 10 & 1.86 \\
       Insulating (intermediate coupling) & 0.16 & 0.83 & 3.5 & 1.28 \\
       Metallic  & 0.16 & 0.83 & 2.5 & 0.74
    \end{tabular}
    \caption{Values for the Hamiltonian parameters (in units of the nearest-neighbor hopping $\tau$) used in this work, and respective staggered magnetic moment per unit cell, in units of Bohr magnetons $\mu_\mathrm{B}$.}
    \label{tab:my_label}
\end{table}

\section{Relationship between the density of Stoner modes and the magnon lifetime}
The standard random phase approximation (RPA) applied to the transverse spin susceptibility
of a Hubbard Hamiltonian results in a relationship between the magnon Green function $\chi^{+-}$ and
the mean-field Green function $\bar{\chi}^{+-}$,
\begin{equation}
    \chi^{+-}(\vec{Q},\hbar\Omega) =  \frac{\bar{\chi}^{+-}(\vec{Q},\hbar\Omega)}{1+U\bar{\chi}^{+-}(\vec{Q},\hbar\Omega)}.
\end{equation}
We would like to cast this expression in a form that resembles a Green function with a self-energy correction,
\begin{equation}
    G = \frac{1}{\bar{G}^{-1}+\Sigma},
\end{equation}
where $\bar{G}$ is the bare Green function and $\Sigma$ is the self-energy. For this it is useful to split all quantities into their real and imaginary parts, denoted below by $R$ and $I$ subscripts. The real and imaginary parts of the magnon Green function then become (we will omit the energy and wave vector arguments for now to avoid cluttering the expressions)
\begin{eqnarray}
    \mathrm{Re}\left[\chi^{+-}\right] =  \frac{\bar{\chi}^{+-}_R(1+U\bar{\chi}^{+-}_R) + U(\bar{\chi}^{+-}_I)^2}{(1+U\bar{\chi}^{+-}_R)^2 + (U\bar{\chi}^{+-}_I)^2}, \nonumber\\
    \mathrm{Im}\left[\chi^{+-}\right] =\frac{\bar{\chi}^{+-}_I}{(1+U\bar{\chi}^{+-}_R)^2 + (U\bar{\chi}^{+-}_I)^2}.
\end{eqnarray}
Similarly,
\begin{eqnarray}
    \mathrm{Re}\left[G\right] =  \frac{\bar{G}^{-1}+\Sigma_R}{(\bar{G}^{-1}+\Sigma_R)^2 + \Sigma_I^2}, \nonumber\\
    \mathrm{Im}\left[G\right] =  -\frac{\Sigma_I}{(\bar{G}^{-1}+\Sigma_R)^2 + \Sigma_I^2}. 
\end{eqnarray}
By comparing the imaginary parts of the generic Green function $G$ to 
$\mathrm{Im}\left[\chi^{+-}\right]$ we notice immediately a clear analogy 
between $U\bar{\chi}^{+-}_I$ and $\Sigma_I$. For small $\hbar\Omega$ and 
fixed wave vector $\vec{Q}$, 
\begin{equation}
    \bar{\chi}^{+-}_I\approx \alpha(\vec{Q})\hbar\Omega,
\end{equation}
where the function $\alpha(\vec{Q})$ depends on the electronic bands in the mean-field configuration. Thus, the magnon's spectral linewidth increases monotonically with energy at small magnon energy and momenta. Notice also that, as in the electronic case, magnon damping is inextricably tied to shifts in magnon energy, through the real part of the self-energy $\Sigma_R$.

\begin{figure}
    \centering
    \includegraphics[width=\columnwidth]{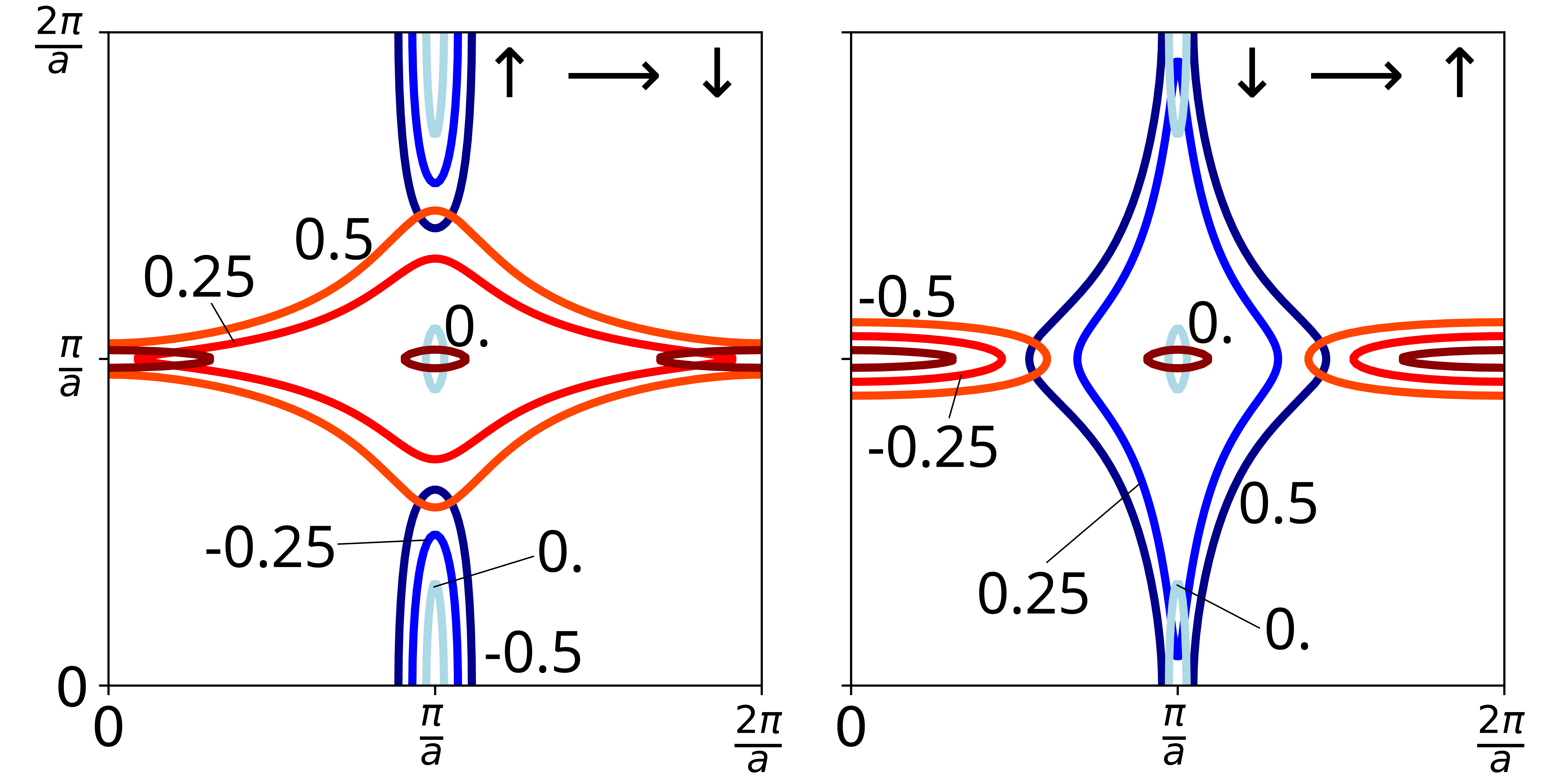}
    \caption{Contours of the electronic bands around the Fermi level; blue curves are for $\up$ spin bands,
    red curves for $\dn$. Left panel ($\up\longrightarrow\dn$): occupied  $\up$ states (shades of blue, at energies $E_F-0.5\tau$, $E_F-0.25\tau$ and $E_F$) and unoccupied $\dn$ states (shades of red, $E_F$, $E_F+0.25\tau$ and $E_F+0.5\tau$). Right panel ($\dn\longrightarrow\up$): occupied  $\dn$ states (shades of red, at energies $E_F-0.5\tau$, $E_F-0.25\tau$ and $E_F$) and unoccupied $\up$ states (shades of blue, $E_F$, $E_F+0.25\tau$ and $E_F+0.5\tau$).}
    \label{fig:energy_contours}
\end{figure}

\section{Origin of the anisotropic altermagnon lifetime}
To further shed light on the mechanism behind the lifetime anisotropy of
metallic altermagnons, it is useful to look at constant energy contours 
of the electronic bands in the mean-field altermagnetic configuration.
The goal is to identify qualitatively the direction dependence of single-particle
spin-flip transitions that give rise to the anisotropic density of 
Stoner modes. In figure~\ref{fig:energy_contours} we show three constant
energy contours for each spin direction, blue contours for $\up$ spin 
electrons, red contours for $\dn$. In the left panel we show contours
for occupied $\up$ states (including the Fermi contour at zero energy) and
unoccupied $\dn$ states (also including the Fermi contour at zero energy), relevant for $S^z=-1$ spin flips ($\dn\longrightarrow\up$).
Thus, in the left panel we can identify possible single-particle spin-flip transitions by connecting blue and red contours. In the left panel we see
that, apart from the very small pockets at $(\frac{\pi}{a},\frac{\pi}{a})$,
there is no horizontal line connecting blue and red contours.
The consequence is that the density of $S^z=-1$ Stoner modes with wave vectors 
along the $x$ direction is very small, and $S^z=-1$ altermagnons propagating 
along the $x$ direction are long-lived. On the other hand, there are plenty of 
connections between blue and red contours at angles $\gtrsim 30^\circ$, meaning
that altermagnons propagating along those directions will be substantially
damped. In the right panel we show the analogous information for $S^z=1$ spin 
flips ($\dn\longrightarrow\up$): occupied $\dn$ states (including the Fermi 
contour at zero energy) and unoccupied $\up$ states (also including the Fermi 
contour at zero energy). Now it is clear that there are many possible single-
particle spin-flip transitions with wave vectors along $x$, whereas very few with
wave vectors along $y$, thus meaning that $S^z=1$ altermagnons are strongly
damped when propagating along $x$ but long-lived when propagating along $x$.

\section{Insulating altermagnet in the intermediate coupling regime ($U=3.5\tau$).} \label{sec:intermediate}

As mentioned in the main main text, the insulating altermagnetic phase of the model is obtained 
for $U\gtrsim 3\tau$. In this regime, although the electronic bands are clearly those of an 
altermagnetic insulator (see the left panel of figure~\ref{fig:el_bands_2}), the altermagnons bear
marks of itinerant magnetism, especially at short wavelengths. A clear signature of itinerant 
behavior is the fact that the magnon lineshape acquires a finite linewidth and, at large enough 
energies, deviates significantly from a a lorentzian shape. This is seen in
fig.~\ref{fig:rho_intermediate} for a short wavelength magnon ($\lambda=2a$) propagating along 
the $x$ direction. The lineshape of the $S^z=1$ magnon (right panel) is very close to a 
lorentzian (dashed orange line). In contrast, the lineshape of the higher energy $S^z=-1$ magnon 
(left panel) is clearly not a lorentzian.

\begin{figure}
    \centering
    \includegraphics[width=\columnwidth]{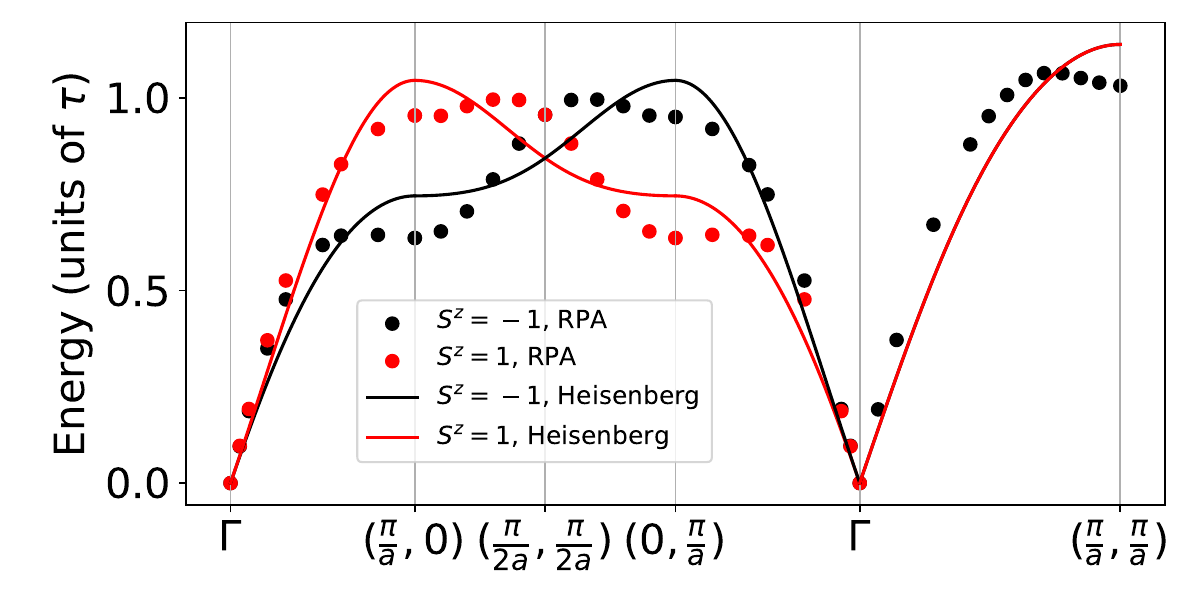}
    \caption{Dispersion relation for magnons in an insulating altermagnet in the intermediate coupling regime ($U=3.5\tau$). The Heisenberg model used to fit the RPA energies includes up to third-neighbor exchange.}
    \label{fig:sw_dispersion_insulating_intermcoupl}
\end{figure}

Another consequence of the coupling between magnons and Stoner excitations is
a renormalization of magnon energies relative to those predicted by a localized 
spin model. In fig.~\ref{fig:sw_dispersion_insulating_intermcoupl} we compare the dispersion relation of magnons for the insulating altermagnet in the intermediate coupling regime, extracted from
the fermionica model, to the energies of linearized Holstein-Primakoff magnons of a localized spins model, with exchanges up to third neighbors. The exchange parameters of the localized spin model have been obtained from a fit to the fermionic model energies. Although the main qualitative features of the dispersion are captured by the localized spins model, it does a poor job of matching quantitatively the magnon energies over the whole
Brillouin zone, since the spin only model cannot capture the renormalization of the magnon energies by Stoner excitations.

 \begin{figure}
    \centering
    \includegraphics[width=\columnwidth]{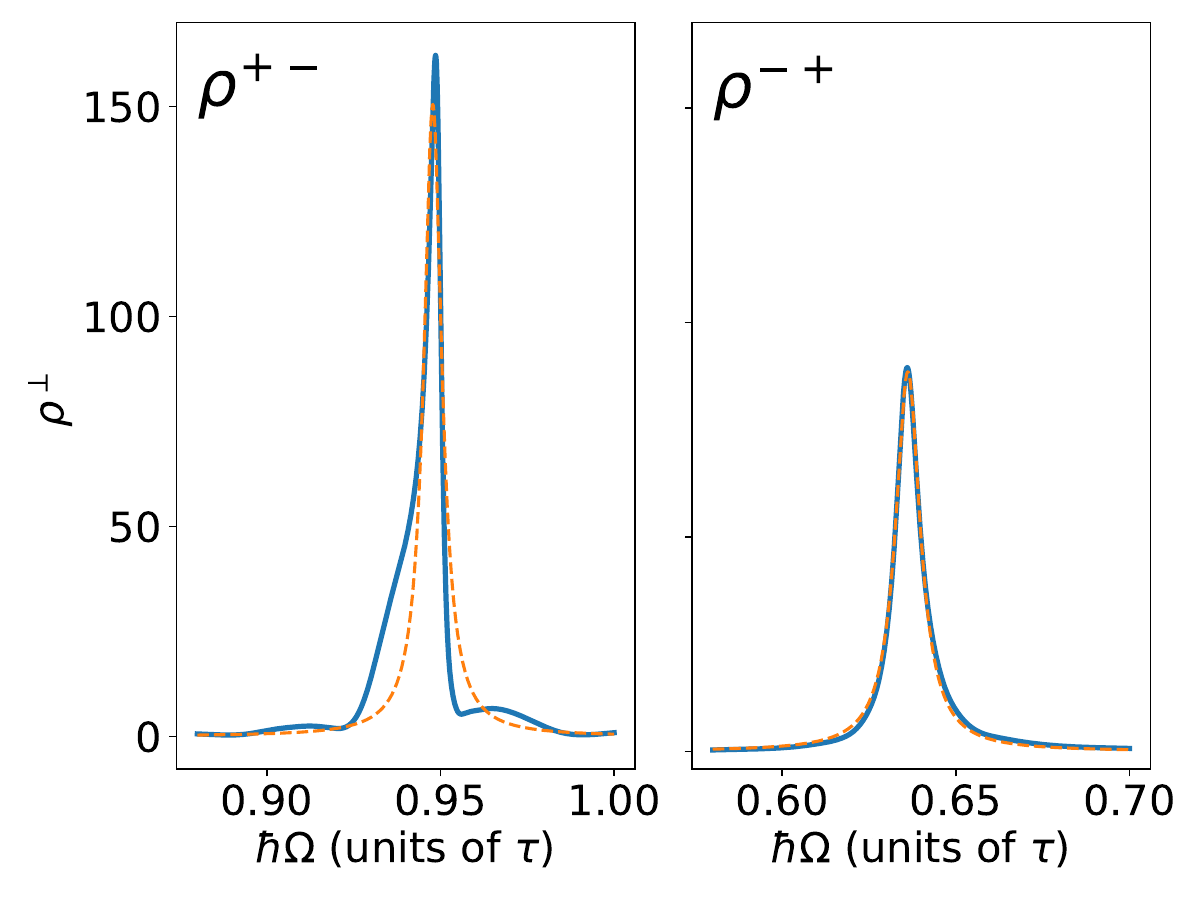}
    \caption{Spectral density of insulating altermagnons in the intermediate
    regime ($U=3.5\tau$), for wave vector $\vec{Q}=(\frac{\pi}{a},0)$. The left and right panels correspond to $S^z=-1$  and $S^z=1$ altermagnons, respectively.}
    \label{fig:rho_intermediate}
\end{figure}

To illustrate the effect of the coupling to the Stoner continuum we plot, in fig.~\ref{fig:rho_intermediate},
the spectral densities for altermagnons with $S^z=-1$ ($\rho^{+-}$) and $S^z=1$ ($\rho^{-+}$). Notice that the
lineshape of the $S^z=-1$ magnon (left panel) is clearly not a lorentzian, whereas the $S^z=1$ magnon
is well fitted by a lorentzian with a finite linewidth, denoting a finite lifetime.

\section{Directionality of the altermagnon spectrum in the insulating regime (intermediate coupling).}

\begin{figure}[h]
    \centering
    \includegraphics[width=\columnwidth]{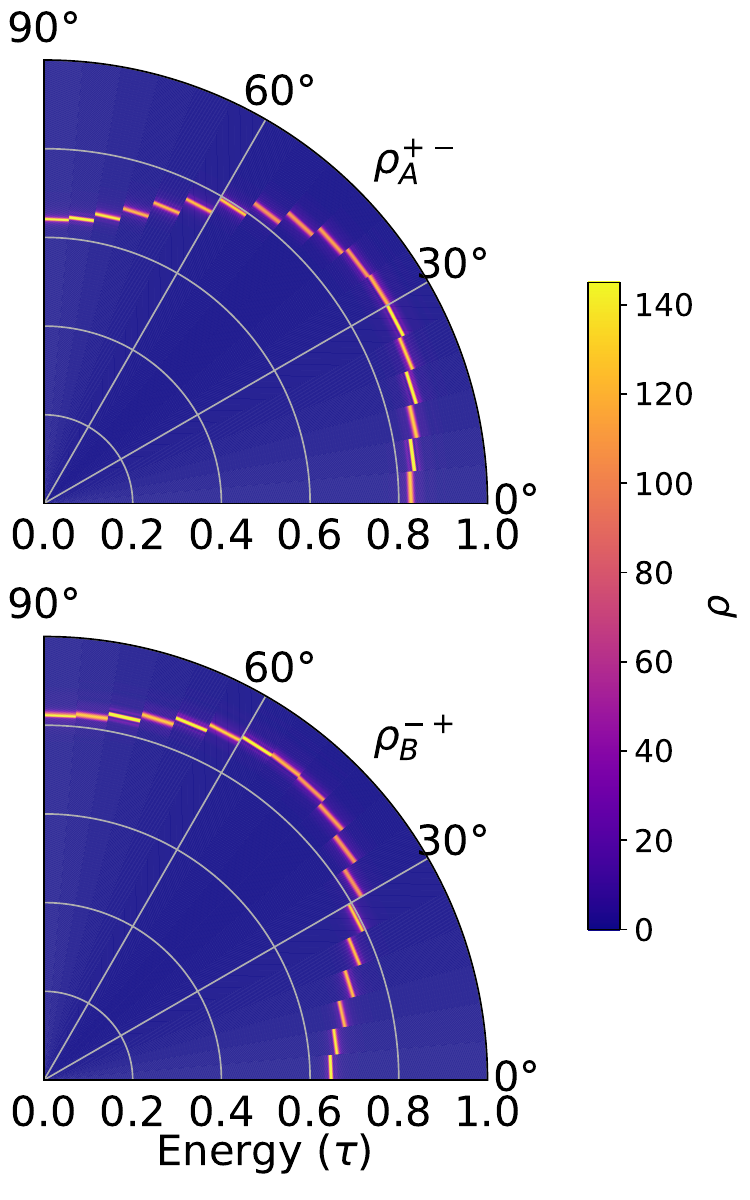}
    \caption{\textbf{Directionality of altermagnons in an insulator}. We plot the altermagnon spectral densities, as a function propagation angle, for a fixed wavelength ($\frac{10a}{3}$) for an insulating altermagnet. The top panel shows $\rho^{+-}_A$ (for the $S^z=-1$ polarization) and the bottom panel shows $\rho^{-+}_B$ (for the $S^z=-1$ polarization). The radial variable represents energy (in units of the nearest-neighbor hopping $\tau$).}
    \label{fig:SM_radial_insulator}
\end{figure}

Here we illustrate the directional dependence of the altermagnon energies for the
intermediate coupling ($U=3.5\tau$) insulating case (figure~\ref{fig:SM_radial_insulator}). The main difference between
this case and the metallic and slightly doped cases is that the altermagnons appear as
well-defined collective excitation for all directions of propagations (compare with fig.~\ref{fig:radialSD} of the main text).

\end{document}